\documentclass[prb,twocolumn,showpacs,amsmath,amssymb,superscriptaddress]{revtex4}

\usepackage{graphicx}
\usepackage{dcolumn}
\usepackage{bm}

\newcommand{\be}{\begin{equation}}
\newcommand{\ee}{\end{equation}}
\newcommand{\bea}{\begin{eqnarray}}
\newcommand{\eea}{\end{eqnarray}}
\newcommand{\mbf}{\mathbf}
\newcommand{\bs}{\boldsymbol}

\begin{document}

\title{Interplay of disorder and magnetic field in the 
  superconducting vortex state}
\author{J. Lages}
\email{lages@cfif.ist.utl.pt}
\author{P. D. Sacramento}
\email{pdss@cfif.ist.utl.pt}
\affiliation{
Centro de F\'\i sica das Interac\c c\~oes Fundamentais,
Instituto Superior T\'ecnico,
Av. Rovisco Pais, 1049-001 Lisboa, Portugal}
\author{Z. Te\v{s}anovi\'c}
\email{zbt@pha.jhu.edu}
\affiliation{
Department of Physics and Astronomy, Johns Hopkins University, Baltimore,
Maryland 21218, USA}

\date{\today}

\begin{abstract}
We calculate the density of states of an inhomogeneous
superconductor in a magnetic field where 
the positions of vortices are distributed completely at random. 
We consider both the cases of $s$-wave
and $d$-wave pairing.
For both pairing symmetries either the presence of disorder
or increasing the density of vortices enhances the low energy density of
states. In the $s$-wave case the gap is filled and the density of 
states is a power law at low energies. In the $d$-wave case the
density of states is finite at zero energy and it rises linearly at
very low energies in the Dirac isotropic case ($\alpha_D=t/\Delta_0=1$, 
where $t$ is the hopping integral 
and $\Delta_0$ is the amplitude of the order parameter). 
For slightly higher energies the density of
states crosses over to a quadratic behavior.
As the Dirac anisotropy increases (as $\Delta_0$
decreases with respect to the hopping term) the linear region decreases
in width. Neglecting this small region the density of states interpolates
between quadratic and back to linear as $\alpha_D$ increases.
The low energy states are strongly peaked near the vortex cores.\end{abstract}


\maketitle

\section{Introduction}

The interaction between the superconductor
quasiparticles and the vortices
induced by an external magnetic field has been a subject of considerable
recent interest \cite{Gorkov,Anderson,FT}. In the presence
of vortices the quasiparticles
feel the combined effect of the external magnetic field
and of the spatially varying
field of the chiral supercurrents. By performing a gauge
transformation to effectively
reduce the system to the one in a zero average magnetic field
it was shown \cite{FT}
that the natural low energy quasiparticle 
modes are Bloch waves rather than the Dirac
Landau levels proposed
in Refs. \onlinecite{Gorkov} and \onlinecite{Anderson}.
The results or Ref. \cite{FT} also showed that the quasiparticles
besides feeling a Doppler shift caused by the moving supercurrents
\cite{Volovik} (Volovik effect)
also feel a quantum ``Berry'' like term due to a half-flux
Aharonov-Bohm scattering
of the quasiparticles by the vortices.

The effect of disorder on the low-energy
density of states of superconductors
has also been a subject of much recent activity \cite{report}, particularly 
in the case of $d$-wave symmetry. 
Several conflicting predictions have appeared 
in the literature which have mainly
concentrated on the effect of the presence of impurities.
Some progress toward understanding the
disparity of theoretical results has been achieved
realising that the details of
the type of disorder affect significantly the density of states \cite{report}.
Particularly in the case of $d$-wave superconductors, in contrast to conventional 
gapped $s$-wave superconductors, the presence of gapless
nodes is expected to affect the transport properties.
Using a field theoretic description and linearizing the spectrum around the 
four Dirac-like nodes it has been suggested
that the system is critical. It was obtained that the density of states is
of the type $\rho(\epsilon) \sim |\epsilon |^{\alpha}$, where $\alpha$
is a non-universal
exponent dependent on the disorder, and that the low
energy modes are extended states
(critical metal) \cite{NTW}. Taking into account the effects of inter-nodal
scattering (hard-scattering) it has been shown that an
insulating state is obtained
instead, where the density of states still vanishes at low energy but with an
exponent $\alpha=1$ independent of disorder \cite{Senthil}.
The addition of time-reversal
breaking creates two new classes designated spin quantum Hall
effect I and II, due
to their similarities to the usual quantum Hall effect,
corresponding to the hard
and soft scattering cases, respectively \cite{report}.
The proposed formation of a
pairing with a symmetry of the type $d+id$ breaks
time-inversion symmetry \cite{did}
but up to now remains a theoretical possibility.
On the other hand applying an external
magnetic field naturally breaks time-reversal invariance and therefore it is
important to study the density of states in this case.

In general, disorder is due to the presence of impurities which may either
scatter the 
quasiparticles and/or
may serve as pinning centers for the field induced vortices.
The density of states of a dirty but homogeneous
$s$-wave superconductor in a high
magnetic field, where the quasiparticles scatter off scalar impurities,
was considered using a Landau level basis \cite{Dukan}. For small amounts of
disorder it was found that $\rho(\epsilon) \sim \epsilon^2$ but when the
disorder is higher than some critical value a finite density of states is
created at the Fermi surface. In the same regime of high magnetic fields, 
but with
randomly pinned vortices and no impurities, the density of states at low 
energies increases
significantly with respect to the lattice case suggesting a finite value at
zero energy \cite{Sacramento}. 
Refs. \onlinecite{Ye} and \onlinecite{Khveshchenko}
considered the effects of random and
statistically independent scalar and vector potentials
on $d$-wave quasiparticles and it was predicted \cite{Khveshchenko} 
that at low energies $\rho(\epsilon) \sim \rho_0 +a \epsilon^2$,
where $\rho_0 \sim B^{1/2}$.
The effect of randomly pinned {\em discrete}
vortices on the spectrum of a $d$-wave superconductor was considered recently 
and
a preliminary report was presented in Ref. \onlinecite{prl} 
for the isotropic case. In this work we study the 
density
of states as a function of vortex density and consider the effects of the Dirac
anisotropy. We also study the local density of states (LDOS) and the inverse 
participation
ratio (IPR)
and study the effect of disorder on the localization of the low energy 
states.

The nature of the low energy states in the presence of a single vortex with 
$s$-wave
symmetry was solved long ago \cite{Caroli}. There are localized bound states 
in the vortex
core. The $d$-wave symmetry case led to some early controversy but it was 
eventually clearly demonstrated
that the low energy states, even though 
strongly peaked near the vortex core, 
extend along the four nodal directions 
and are indeed delocalized as shown by the 
behavior of IPR \cite{dvortex}. 
In the vortex lattice the states are 
naturally also extended \cite{FT,PRB}.
In the clean limit it is therefore 
expected that the external field will increase the 
low-energy
density of states. The expectation that 
these are delocalized is evidenced by the 
increase
of the thermal conductivity with field \cite{Chiao1} 
in contrast to the reducing
effect of conventional superconductors \cite{Bessa}.

The addition of impurities in zero magnetic field 
has been studied using the Bogoliubov-de Gennes (BdG) equations. It was found 
that the
$d$-wave superconductivity is mainly destroyed locally near a strong scatterer.
The superfluid density is strongly suppressed near the impurities but 
only mildly
affected elsewhere \cite{Atkinson}. No evidence for localization of the 
low energy
states was found and accordingly the superfluid density is indeed suppressed
but less than expected \cite{Franzet,Randeria1} and, accordingly, the decrease 
of the
critical temperature with disorder is much slower than previously expected in
accordance with experiments \cite{Ulm}. Similar results of an inhomogeneous 
order parameter
were also obtained for $s$-wave superconductors \cite{Randeria2}.

The question we address in this paper is the influence of the positional 
disorder of the vortices on the quasiparticle states of either a
$s$- or a $d$-wave superconductor in an external magnetic field and we will
not consider the scattering off impurities.
We obtain that the low energy states are strongly
peaked near the vortex locations as evidenced by the LDOS.
At higher energies the states have a very uniform distribution throughout the 
system.
For these states the IPR scales as $1/L^2$ indicative 
of
extended states. In the case of the low energy states the same quantity does 
not follow
this scaling but does not saturate either. This is probably a consequence 
of finite
size effects indicative that the system sizes considered are smaller than the
localization length. The results indicate at best
a large value for the localization 
length
but are more consistent with extended states rather than with localized 
states.

In section II we describe the model and present the BdG equations to be solved
and consider the effect of the vortices on the supercurrent profiles.
In section III we study the effects of positional disorder on s-wave superconductors.
In section IV we consider d-wave superconductors for the isotropic and anisotropic cases.
Also we study the spatial structure of the low-lying quasiparticle states. We end
with the conclusions.

\section{Description of the model}

The strong correlations of the high $T_c$ materials are typically modelled
 using
a Hubbard-like Hamiltonian for the electrons. The superconductivity is 
considered
using a BCS-like formulation which provides good agreement with experimental 
results.
Even though the normal phase of these materials is particularly challenging
it is by now accepted that the phenomenology in the superconducting phase is
reasonably well described by BCS theory.

\subsection{Bogoliubov-de Gennes Hamiltonian}

We will consider the lattice
formulation of a disordered $d$-wave superconductor
in a magnetic field. We start from the Bogoliubov-de Gennes equations
${\cal H} \psi = \epsilon \psi$
where $\psi^{\dagger}(\mbf{r})=\left(u^*(\mbf{r}),
v^*(\mbf{r}) \right)$ and where the matrix Hamiltonian is given by
\begin{equation} \label{H}
{\cal H} = \left( \begin{array}{cc}
\hat{h} & \hat{\Delta} \\
\hat{\Delta}^{\dagger} & -\hat{h}^{\dagger} \\
\end{array} \right)
\end{equation}
with \cite{FT,PRB}
\begin{equation}\label{h}
\hat{h} = -t \sum_{\bs{\delta}} e^{-\frac{ie}{\hbar c}
\int_{\mbf{r}}^{\mbf{r}+
\mbf{\bs{\delta}}} \mbf{A}(\mbf{r}) \cdot
d\mbf{l}}
\hat{s}_{\bs{\delta}} - \epsilon_F
\end{equation}
and
\begin{equation}\label{delta}
\hat{\Delta} = \Delta_0 \sum_{\bs{\delta}} e^{\frac{i}{2}
\phi(\mbf{r})}
\hat{\eta}_{\bs{\delta}} e^{\frac{i}{2} \phi(\mbf{r})}.
\end{equation}
The sums are over nearest neighbors ($\bs{\delta}=\pm
\mbf{x}, \pm \mbf{y}$
on the square lattice); $\mbf{A}(\mbf{r})$ is the vector
potential associated with the uniform external magnetic field
$\mbf{B}=\bs{\nabla}\times\mbf{A}$,
the operator $\hat{s}_{\bs{\delta}}$ is defined through
its action on space dependent functions,
$\hat{s}_{\bs{\delta}}u(\mbf{r}) = u(\mbf{r}+\bs{\delta})$, and
the operator $\hat{\eta}_{\bs{\delta}}$ describes the symmetry
of the order parameter.

The application of the uniform external magnetic field $\mbf{B}$
generates in type II superconductors compensating vortices each
carying one half of the magnetic quantum flux,
$\displaystyle\frac{\phi_0}{2}$.
We take the London limit, which is valid for low magnetic field
and over most of the $H-T$ phase diagram in extreme type-II
superconductors like the cuprates,
assuming that the size of the vortex cores is
negligible and placing each vortex core at the center of
a plaquette (unit cell). The $N_{\phi}$ vortices are distributed
randomly over the $L \times L$ plaquettes.
In this limit we can take the order parameter
amplitude $\Delta_0$ constant everywhere in space and we can
factorize the phase of the order parameter as shown in Eq. \ref{delta}.

\begin{figure}
\includegraphics[width=7cm,height=7cm]{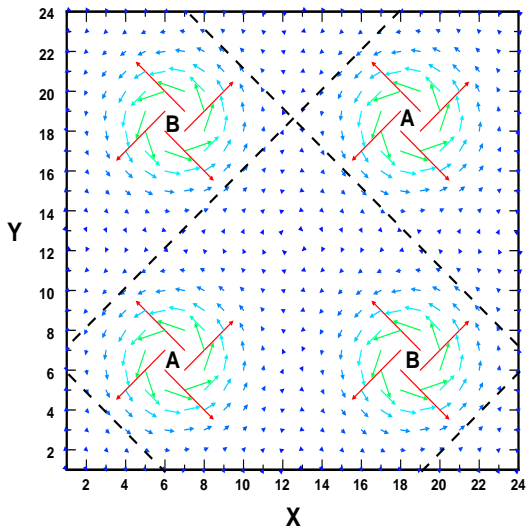}
\caption{\label{fig1}
Color vector field plot showing the profile of the conventional
superfluid wave vector $\mbf{k}_s=
\frac{1}{2}\left(\mbf{k}_s^A+\mbf{k}_s^B\right)$ for a regular vortex lattice
and for $B=1/144$.
}
\end{figure}

At this stage, it is convenient to perform a singular gauge
transformation to eliminate the phase of the
off-diagonal term (\ref{delta}) in the matrix Hamiltonian.
We consider the unitary FT gauge transformation 
${\cal H} \rightarrow {\cal U}^{-1} {\cal H} U$, where\cite{FT}
\begin{equation} \label{gauge}
{\cal U} = \left( \begin{array}{cc}
e^{i \phi_A(\mbf{r})} & 0 \\
0 & e^{-i \phi_B(\mbf{r})} \\
\end{array} \right)
\end{equation}
with $\phi_A(\mbf{r})+\phi_B(\mbf{r})=\phi(\mbf{r})$.
The phase field $\phi(\mbf{r})$ is then decomposed at each site
of the two-dimensional lattice in two components $\phi_A(\mbf{r})$
and $\phi_B(\mbf{r})$ which are assigned respectively 
to a set of vortices $A$, positioned at $\{\mbf{r}_i^A \}_{i=1,N_A}$,
and a set of vortices $B$, positioned at $\{\mbf{r}_i^B \}_{i=1,N_B}$.
The phase fields $\phi_{\mu=A,B}$ are naturally defined through the equation
\begin{equation}
\bs{\nabla} \times \bs{\nabla} \phi_{\mu}
(\mbf{r}) = 2 \pi \mbf{z}
\sum_i \delta (\mbf{r}-\mbf{r}_i^{\mu})
\end{equation}
where the sum runs only over the $\mu$-type vortices.
After carying out the gauge transformation (\ref{gauge})
the Hamiltonian (\ref{H}) reads
\begin{equation}\label{Hp}
\mathcal{H}'=
\displaystyle
\left(
\begin{array}{cc}
-t\displaystyle\sum_{\bs{\delta}}
e^{i\mathcal{V}^A_{\bs{\delta}}\left(\mbf{r}\right)}\hat{s}_{\bs{\delta}}
-\epsilon_F
&
\Delta_0\displaystyle\sum_{\bs{\delta}}
e^{-i\frac{\delta\phi}{2}}
\hat\eta_{\bs{\delta}}
e^{i\frac{\delta\phi}{2}}
\\
\Delta_0\displaystyle\sum_{\bs{\delta}}
e^{-i\frac{\delta\phi}{2}}
\hat\eta^\dagger_{\bs{\delta}}
e^{i\frac{\delta\phi}{2}}
&
t\displaystyle\sum_{\bs{\delta}}
e^{-i\mathcal{V}^B_{\bs{\delta}}\left(\mbf{r}\right)}\hat{s}_{\bs{\delta}}
+\epsilon_F
\end{array}
\right).
\end{equation}
The phase factors are given by \cite{PRB}
${\cal V}_{\bs{\delta}}^{\mu}(\mbf{r})
=\int_{\mbf{r}}^{\mbf{r}+\bs{\delta}}
\mbf{k}_s^{\mu}
\cdot d\mbf{l}$ and 
$\delta\phi(\mbf{r})=\phi_A(\mbf{r})-\phi_B(\mbf{r})$,
where $\hbar\mbf{k}_s^{\mu} = m \mbf{v}_s^{\mu} =
\hbar\bs{\nabla}
\phi_{\mu} -
\frac{e}{c} \mbf{A}$ is the superfluid momentum 
vector for the $\mu$-supercurrent.
Physically, the vortices $A$ are only
visible to the particles and the vortices $B$ are only visible to the holes.
Each resulting $\mu$-subsystem is then in an effective magnetic field
\begin{equation}\label{Bmu}
\mbf{B}_{\mathrm{eff}}^\mu=-\frac{mc}{e}\bs{\nabla}\times\mbf{v}^\mu_s
=\mbf{B}-\phi_0\mbf{z}\sum_i\delta^2(\mbf{r}-\mbf{r}_i^\mu)
\end{equation}
where each vortex carries now an effective quantum magnetic flux $\phi_0$.
For the case of a regular vortex lattice\cite{FT,PRB}, these effective magnetic
fields vanish simultaneously on average if the magnetic unit cell
contains two vortices, one of each type.
More generally, in the absence of spatial symmetries, as it is the case
for disordered systems, these effective
magnetic fields $\mbf{B}_{\mathrm{eff}}^{\mu=A,B}$ vanish if
the numbers of vortices of the two types $A$ and $B$
are equal, i.e. $N_A=N_B$,
and their sum equals to the number of elementary 
quantum fluxes of the external
magnetic field penetrating the system. 
As the number of vortices $N_\phi$
in the two-dimensional system of size $L\times L$
is proportional to the quantized magnetic flux piercing through the
system, we choose to parametrize
the magnetic field intensity by the ratio of the number of
vortices, $N_\phi=N_A+N_B$, by the number of lattice sites, 
$B=\frac{N_\phi}{L\times L}$.

\begin{figure}
\includegraphics[width=7cm,height=7cm]{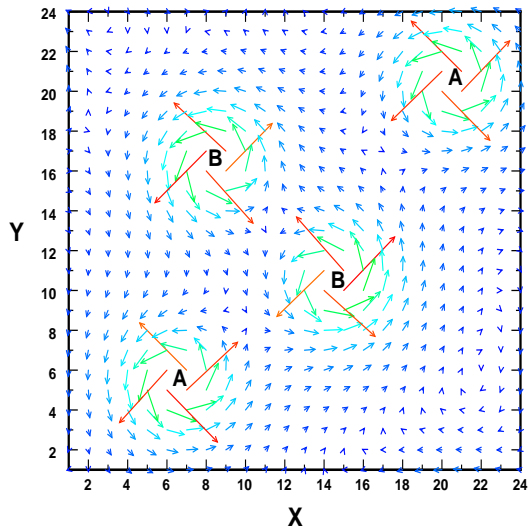}
\caption{\label{fig2}
Color vector field plot showing the profile of the conventional
superfluid wave vector $\mbf{k}_s=
\frac{1}{2}\left(\mbf{k}_s^A+\mbf{k}_s^B\right)$ for a random configuration
of vortices and for $B=1/144$.
}
\end{figure}

We consider here the disorder induced by the random pinning
of the $N_\phi$ vortices over the two-dimensional $L\times L$
lattice. As the effective magnetic fields (\ref{Bmu}) experienced by the
particles and the holes vanish on average within the gauge transformation (\ref{gauge})
we are allowed to use periodic boundary conditions on the square lattice
($\Psi(x+nL,y)=\Psi(x,y+mL)=\Psi(x,y)$ with $n,m\in \Bbb{Z}$). The
$L\times L$ original lattice becomes
then a magnetic supercell where the vortices
are placed at random with a mean intervortex spacing $\ell=1/\sqrt{B}$.
The disorder in the system is then established over a length $L$.

In order to compute the eigenvalues and eigenvectors
of the Hamiltonian (\ref{Hp}) we seek for eigensolutions in the Bloch
form
$\Psi^\dagger_{n\mbf{k}}(\mbf{r})=e^{-i\mbf{k}\cdot\mbf{r}}
(U^*_{n\mbf{k}},V^*_{n\mbf{k}})$ where $\mbf{k}$ is a point of the
Brillouin zone. 
In the following we will label these eigensolutions with the index
$\mbf{n}=(n,\mbf{k})$.
We diagonalize then the Hamiltonian
$e^{-i\mbf{k}\cdot\mbf{r}}\mathcal{H}'e^{i\mbf{k}\cdot\mbf{r}}$
for a large number of points $\mbf{k}$ in the Brillouin zone
and for many different
realizations (around 100) of the random vortex pinning.
The results of these computations are shown in section \ref{swave} 
for the $s$-wave disordered superconductor case
and
in section \ref{dwave} for the $d$-wave disordered superconductor case.

\subsection{Profiles of supercurrents}

\begin{figure}
\includegraphics[width=7cm,height=7cm]{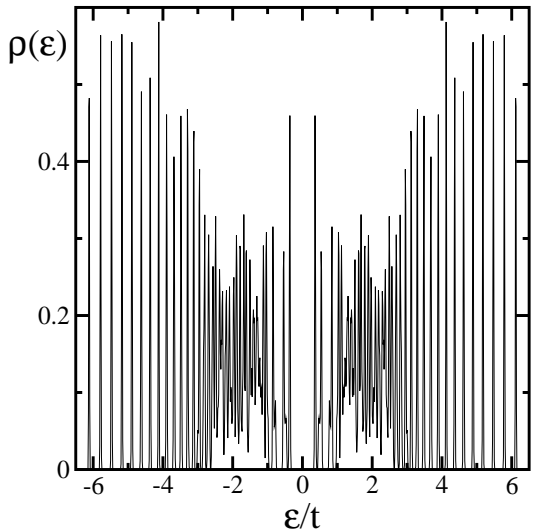}
\caption{\label{fig3}
Quasiparticle density of states for the $s$-wave symmetry and for a
regular lattice of vortices (no disorder). The magnetic field
is $B=1/18$, the magnetic unit cell contains 2 vortices piercing an area
of $6\times 6$.
The parameters are $\mu=-2.2t$ and $\Delta_0=t$.
}
\end{figure}

The $\mu$-superfluid wave vector $\mbf{k}_s^{\mu}(\mbf{r})$
characterizes the supercurrents
induced by the $\mu$-vortices. 
This
vector can be calculated for an arbitrary
configuration of vortices\cite{PRB} like
\be\label{kappamu}
\mbf{k}_s^{\mu}(\mbf{r})=2 \pi \int
\frac{d^2k}{(2 \pi)^2}
\frac{i \mbf{k} \times \mbf{z}}{ k^2+\lambda^{-2}}
\sum_{i=1}^\infty e^{i\mbf{k} \cdot (\mbf{r}-
\mbf{r}_i^{\mu})}.
\ee
Here $\lambda$ is the magnetic penetration length and the sum extends over
the infinite number of $\mu$-type vortex
positions.
We consider the case $\lambda \rightarrow \infty$ which is
an excellent approximation in extreme type-II systems like
high temperature superconductors.
In this limit the repulsive interaction between vortices is not screened
and therefore the vortex distribution is not strictly arbitrary -- long
range interactions will try to force the vortex system to take only
incompressible configurations. For the purposes of this paper,
we assume that the pinning centers are strong enough to overcome 
the vortex repulsion over the lengthscales relevant to experiments.
Thus, we are considering the limit of strong pinning.
We have checked that the case with a random distribution of vortices
is qualitatively the same as for the case where the vortex
positions are allowed to vary with a radius of a few unit cells around a
regular lattice position.

The $\mu$-superfluid wave vector distribution is completely determined
by the configuration of the $\mu$-vortices, as can be seen from 
Eq. \ref{kappamu}
and is independent of the pairing symmetry. Since we are 
taking the London limit,
where the vortex core size is negligible, we neglect any possible different
symmetry contributions from inside the vortex core \cite{Inside} and the chiral
supercurrents simply reflect the circulation of the superfluid density
around the vortex.

In order to compute efficiently the $\mu$-superfluid wave vector
we use the translational symmetry introduced by the periodic repetition of
the $L\times L$ supercells and then we are able to use the Fourier series
representation of expression (\ref{kappamu}), e.g.
\begin{equation}
\mbf{k}_s^{\mu}(\mbf{r})=\frac{2i\pi}{(L\bs{\delta})^2}
\sum_{\bs{Q}\neq \mathbf{0}}
\frac{\mbf{Q}\times\mbf{z}}{\mbf{Q}^2}
\sum_{i=1}^{N_\mu}
e^{i\mbf{Q}\cdot\left(\mbf{r}-\mbf{r}_i^\mu\right)}
\end{equation}
where $\mbf{Q}=\frac{2\pi}{L\delta}\left(n_1,n_2\right)$ with $n_1,n_2\in\Bbb{Z}$.
We remark that the sum over vortex positions 
runs now only over the $\mu$-type vortices
within a supercell and not, as in Eq. \ref{kappamu},
over the infinite number of $\mu$-type
vortices present in the two-dimensional system. The price to pay
for this procedure is the preceding infinite sum over the 
reciprocal vectors $\mbf{Q}$. We truncate this latter sum when the
convergence of each components
of $\bf{k}_s^{\mu}(\mbf{r})$ in each lattice point $\mbf{r}$ is attained.

\begin{figure}
\includegraphics[width=7cm,height=7cm]{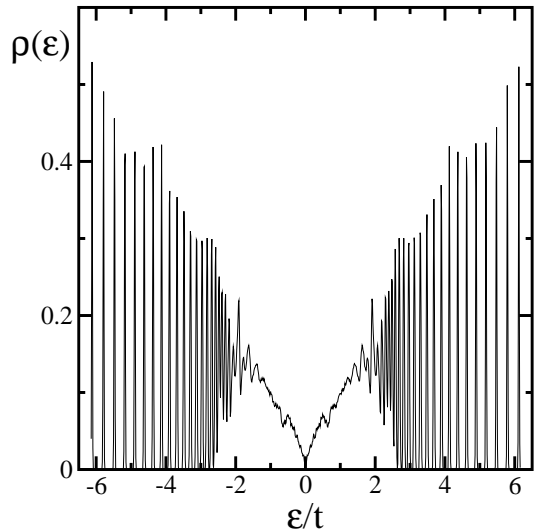}
\caption{\label{fig4}
Quasiparticle density of states for the $s$-wave symmetry and for a
disordered lattice of vortices. The magnetic field
is $B=1/18$ and the linear system size is $L=18$.
The parameters are $\mu=-2.2t$ and $\Delta_0=t$.
}
\end{figure}

Figures \ref{fig1} and \ref{fig2}
show examples of distributions of the conventional superfluid
wave vector which is half the sum of the two types of superfluid
wave vector,
\begin{equation}
\mbf{k}_s(\mbf{r})=\displaystyle
\frac{\mbf{k}_s^{A}(\mbf{r})+\mbf{k}_s^{B}(\mbf{r})}{2}
=
\frac 12\bs{\nabla}\phi -
\frac{e}{c} \mbf{A}.
\end{equation}
In Fig. \ref{fig1} we show the profile of the supercurrent velocities in 
the lattice case.
The unit cell has two vortices, one of each type $A$ and $B$.
In Figure 2 we show the supercurrents for
an arbitrary configuration of the vortices.
For example, the $24\times 24$ lattice shown in Fig. \ref{fig2}
can illustrate a disordered supercell corresponding to a magnetic field
$B=1/144$ with two $A$-type vortices and two $B$-type vortices.
As it should be,
the labelling of the $A$- and the $B$-type vortices is completely
arbitrary.
The intensity of the 
supervelocities for the regular and the disordered vortex lattice
peak around each vortex and decrease rapidly outside the vicinity
of the vortex core ($\sim 5\delta$).

\section{Disordered $s$-wave superconductors}
\label{swave}

\begin{figure}
\includegraphics[width=6.5cm,height=9cm]{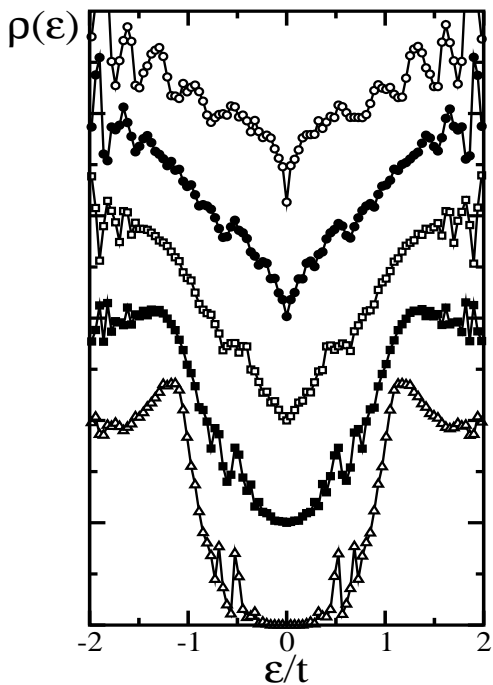}
\caption{\label{fig5}
Quasiparticle density of states ($s$-wave) for different
magnetic fields
$B=2/200$ ($\triangle$),
$B=4/200$ ($\blacksquare$),
$B=7/200$ ($\square$),
$B=11/200$ ($\bullet$)
and
$B=20/200$ ($\circ$)
in units of $hc/(2e\delta^2)$.
The linear system size is $L=20$ and the parameters are
$\mu=-2.2t$, $\Delta_0=t$.
For clarity the different curves are vertically shifted.
}
\end{figure}

For the conventionnal $s$-wave case the operator characterizing
the symmetry of the order parameter is constant
$\hat\eta_{\mbf{\delta}}=\frac 14$ and the off-diagonal terms 
of the Hamiltonian (\ref{Hp})
are then considerably simplified
\begin{equation}\label{Hs}
\mathcal{H}'=
\displaystyle
\left(
\begin{array}{cc}
-t\displaystyle\sum_{\bs{\delta}}
e^{i\mathcal{V}^A_{\bs{\delta}}\left(\mbf{r}\right)}\hat{s}_{\bs{\delta}}
-\epsilon_F
&
\Delta_0
\\
\Delta_0
&
t\displaystyle\sum_{\bs{\delta}}
e^{-i\mathcal{V}^B_{\bs{\delta}}\left(\mbf{r}\right)}\hat{s}_{\bs{\delta}}
+\epsilon_F
\end{array}
\right).
\end{equation}

The $s$-wave Hamiltonian (\ref{Hs}) describes coupled particles and holes
evolving each in an effective magnetic field composed by the external
magnetic field and the counteracting field of $\phi_0$ flux carying vortices.
The effective magnetic fields are characterized by the Peierls phase factors
$\mathcal{V}_{\bs{\delta}}^{\mu=A,B}$. Independently of the system
being disordered or not, the tails of the quasiparticles spectrum 
($\epsilon\gg\Delta_0$) are expected to be described by quantized Landau
levels.

In Fig. 3 we show the density of states for a field $B=1/18$ in the case
of no disorder.
The gap is clearly seen at low energies characteristic of $s$-wave symmetry.
The states inside the gap (for $\arrowvert\epsilon\arrowvert<t$ in Fig.
\ref{fig3}) are the typical 
Caroli - de Gennes - Matricon (CdGM) bound states\cite{Caroli}.
At higher
energies the Landau levels are clearly visible.

\begin{figure}
\includegraphics[width=7cm,height=7cm]{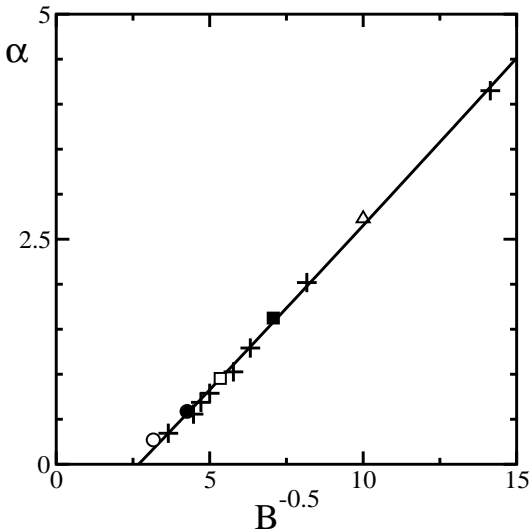}
\caption{\label{fig6}
Magnetic field dependence of the exponent $\alpha$
extracted from the power-law fit $\rho(\epsilon)\sim\epsilon^\alpha$
of the data presented in Fig. \ref{fig5}. The symbols ($\triangle$),
($\blacksquare$), ($\square$), ($\bullet$) and ($\circ$)
denote the same values of $B$ as in Fig.\ref{fig5}; the symbols
($+$) denote others values of $B$ not presented in Fig.\ref{fig5}.
}
\end{figure}

In Fig. 4 we show the
effects of disorder
for the same $B$ field in a lattice of $18 \times 18$ sites ($18$ vortices).
The gap is filled by the disorder. The reader should
observe here that our disorder is in certain sense ``infinite''
since we place vortex positions completely at random. Thus,
it appears that such full randomness in vortex positions, generated
by strong pinning, suffices to close the gap in the single
particle density of states.
In constrast, the Landau level quantization structure clearly
persists at high excitation energies.
In general the increase of the magnetic field modifies the curvature of the 
quasiparticle density of states for the low energies.
As it is shown in Fig. \ref{fig5} the density of states seems
to be decribed by the power-law formula
\begin{equation}
\rho(\epsilon)\sim \epsilon^\alpha.
\end{equation}
In Fig. \ref{fig6} we fit the magnetic field dependence
of the exponent $\alpha$. This exponent obeys the following
law
\begin{equation}
\alpha\simeq c\ell-d
\end{equation}
where $\ell=1/\sqrt{B}$
is the mean intervortex spacing and $d$ is found close to $1$.
In the cases of a strong magnetic field the density of vortices is high and
strictly we are in a regime where the size of the vortex cores can not be
neglected. In this regime the Landau level structure 
at high energies is clear and it extends to low energies superimposed by the
effects of disorder. The lower limit for this high magnetic field 
regime is the value $B\simeq 0.16$ for which $\alpha=0$ in Fig. \ref{fig6}.

\section{Disordered $d$-wave superconductors}
\label{dwave}

For the unconventional $d$-wave case the operator $\hat\eta_{\bs{\delta}}$
takes the form $\hat\eta_{\bs{\delta}}=(-1)^{\delta_y}\hat{s}_{\bs{\delta}}$,
we recall that $\hat{s}_{\bs{\delta}}$ acts on spatial
dependent functions as
$\hat{s}_{\bs{\delta}}u(\mbf{r})=u(\mbf{r}+\bs{\delta})$
and that 
$\bs{\delta}=\pm\mbf{x},\pm\mbf{y}$ characterizes unit displacements (hops)
on the lattice. With these definitions the $d$-wave Hamiltonian can
be derived from the Hamiltonian (\ref{Hp}) and reads
\begin{equation}\label{Hd}
\mathcal{H}'=
\displaystyle
\left(
\begin{array}{cc}
-t\displaystyle\sum_{\bs{\delta}}
e^{i\mathcal{V}^A_{\bs{\delta}}\left(\mbf{r}\right)}\hat{s}_{\bs{\delta}}
-\epsilon_F
&
\Delta_0\displaystyle\sum_{\bs{\delta}}
e^{i\mathcal{A}_{\bs{\delta}}(\mbf{r})+i\pi\delta_y}
\hat{s}_{\bs{\delta}}
\\
\Delta_0\displaystyle\sum_{\bs{\delta}}
e^{-i\mathcal{A}_{\bs{\delta}}(\mbf{r})-i\pi\delta_y}
\hat{s}_{\bs{\delta}}
&
t\displaystyle\sum_{\bs{\delta}}
e^{-i\mathcal{V}^B_{\bs{\delta}}\left(\mbf{r}\right)}\hat{s}_{\bs{\delta}}
+\epsilon_F
\end{array}
\right)
\end{equation}
where the phase factor $\mathcal{A}_{\bs{\delta}}(\mbf{r})$ has the form
\begin{eqnarray}
\mathcal{A}_{\bs{\delta}}(\mbf{r})&=&\frac{1}{2}
\int_{\mbf{r}}^{\mbf{r}+\bs{\delta}}\nonumber
\left(\bs{\nabla}\phi_A-\bs{\nabla}\phi_B\right)\cdot d\mbf{l}\\
&=&\frac{1}{2}\label{Adelta}
\int_{\mbf{r}}^{\mbf{r}+\bs{\delta}}
\left(\mbf{k}_s^A-\mbf{k}_s^B\right)\cdot d\mbf{l}.
\end{eqnarray}
In the Hamiltonian (\ref{Hd}) and in Eq. \ref{Adelta} the vector
\begin{equation}
\mbf{a}_s=\frac 12\left(\mbf{k}_s^A-\mbf{k}_s^B\right)
\end{equation}
acts as an internal gauge field independent of the external magnetic
field\cite{PRB}. The associated internal magnetic field
$\mbf{b}=\bs{\nabla}\times\mbf{a}_s$ consists of opposite
$A-B$ spikes fluxes carying each one half of the magnetic quantum flux
$\phi_0$, centered in the vortex cores and vanishing on average
since the numbers of $A$- and $B$-type vortices are the same.

\subsection{Differences from the regular vortex lattice case}

\begin{figure}
\includegraphics[width=7cm,height=7cm]{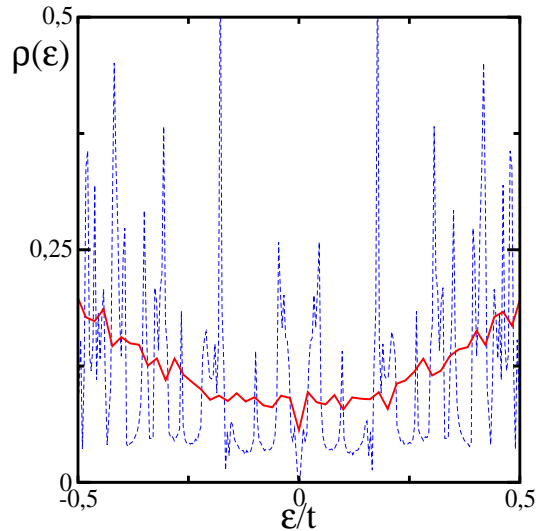}
\caption{\label{fig7}
Quasiparticle density of states for the $d$-wave case
without (dashed line) and with disorder (solid line).
The intensity of the magnetic field is $B=1/121$, $\Delta=0.5t$, $\mu=-2.2t$
and the linear system size is $L=22$.
}
\end{figure}

The situation where the vortices are 
regularly distributed in a lattice was
treated before \cite{PRB}. Since the average 
effective magnetic field vanishes it is possible
to solve the BdG equations using a standard Bloch basis-- the
supercurrent velocities are periodic in space and there is no need to consider
the magnetic Brillouin zone. Taking the continuum limit and linearizing the
spectrum around each node effectively decouples the nodes.
It was shown that the low-energy
quasiparticles are then naturally described as Bloch waves \cite{FT}
and not Dirac-Landau levels as
previously proposed \cite{Gorkov,Anderson}. However, it was found
that in the linearized problem different
assignments of the $A$ and $B$ vortices lead to
somewhat different spectra, which was
unexpected \cite{PRB}. It was found that defining
theory on the lattice regularized this problem and
indeed the system has a manifest internal
gauge symmetry such that the spectrum is independent
of the $A$-$B$ vortex assignments, as
it should be. Moreover, the lattice formulation
explicitly involves internodal
contributions which are
important for the properties of the
density of states in the disordered case.
In the vortex lattice case, however, it was
found that only in special commensurate
cases (for the square lattice) the inclusion of the internodal contributions
is relevant since only in such cases a gap
develops due to the interference
terms between the various nodes, estimated to be of
the order of $\sqrt{B}$. In a general
incommensurate case the interference
is not relevant leading to qualitatively
similar spectra. In the $d$-wave case the spectrum is gapless with a linear
density of states at low energy \cite{PRB}. One would therefore
expect that in a general disordered vortex case
internodal scattering might not be relevant (particularly for
high Dirac cone anisotropy $\alpha_D=v_F/v_\Delta=t/\Delta_0\gg 1$).

\begin{figure}
\includegraphics[width=7cm,height=7cm]{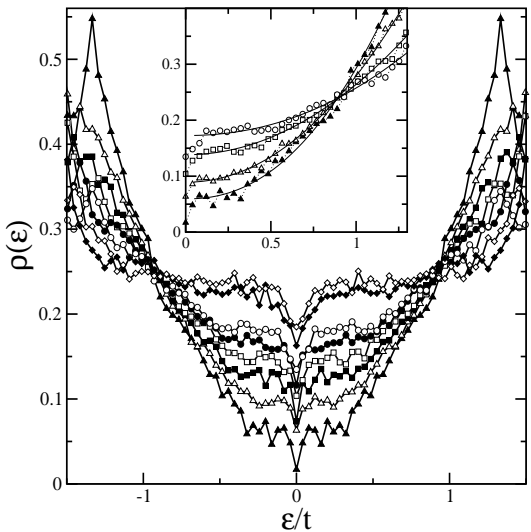}
\caption{\label{fig8}
Quasiparticle density of states for the $d$-wave case and for different
magnetic fields
$B=1/200$ ($\blacktriangle$),
$B=3/200$ ($\triangle$),
$B=5/200$ ($\blacksquare$),
$B=7/200$ ($\square$),
$B=9/200$ ($\bullet$),
$B=11/200$ ($\circ$),
$B=20/200$ ($\blacklozenge$)
and
$B=25/200$ ($\lozenge$)
in units of $hc/(2e\delta^2)$.
The linear system size is $L=20$ and the parameters are
$\mu=-2.2t$, $\Delta_0=t$.
The inset shows the fits of the density of states (solid lines)
inside the presented energy interval. For clarity not all the fits are shown.
}
\end{figure}

In Fig. \ref{fig7}
we compare the densities of states for the lattice case and a case
with disorder.
At weak fields the density of states is small at
low energies having a dip close to zero
energy. We have checked for finite size effects on the spectrum. For system
sizes larger than $16 \times 16$ the density of states at not very low energies
converges and the finite size dependence is negligible. 

\subsection{Isotropic case}

In this subsection we will focus our attention on the isotropic 
case $\alpha_D=1$ in order to extract the general dependencies 
of the quasiparticle density of states in the magnetic field $B$.

\begin{figure}
\includegraphics[width=7cm,height=7cm]{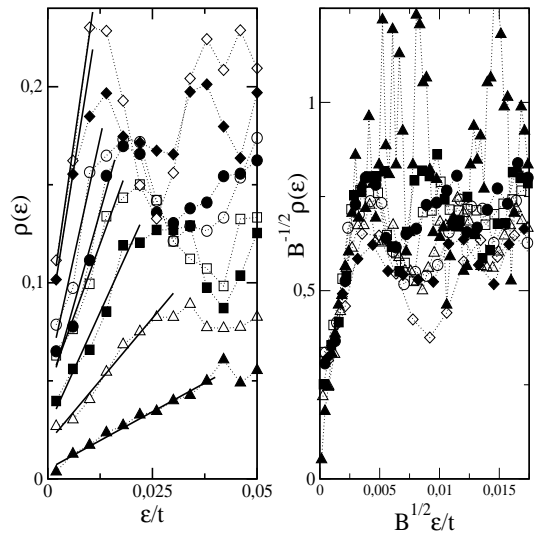}
\caption{\label{fig9} Low-energy quasiparticle density of states
$\rho(\epsilon)$ for the $d$-wave case and for different
magnetic field
$B=1/200$ ($\blacktriangle$),
$B=3/200$ ($\triangle$),
$B=5/200$ ($\blacksquare$),
$B=7/200$ ($\square$),
$B=9/200$ ($\bullet$),
$B=11/200$ ($\circ$),
$B=20/200$ ($\blacklozenge$)
and
$B=25/200$ ($\lozenge$)
in unit of $hc/(2e\delta^2)$.
For each data sets the linear system size is $L=20$,
the same as in Fig. \ref{fig8}. In the left panel the solid lines are linear fits
of the dip region below $\epsilon=0.02t$
of the type $\rho (\epsilon) = \rho_{0\rm{dip}}+ \beta | \epsilon |$. In the
right panel we present the near scaling at low energies (see text).}
\end{figure}

In Fig. \ref{fig8} we show the density of
states for a system with size $20 \times 20$
and for various magnetic fields. The density
of states at small energies is finite
up to quite small energies where there is a
dip to a value that decreases as the
magnetic field decreases. Only for quite small
magnetic fields the density of states
approaches zero at the origin. Neglecting the
narrow region close to the origin we have
fitted the density of states using the power law
\be\label{powerlaw}
\rho (\epsilon) = \rho_0 + \beta | \epsilon|^{\alpha}.
\ee
In the inset of Fig. \ref{fig8}
we show the fits for the various values of the
magnetic field. Reasonable fits are obtained taking $\alpha \sim 2$ and
we obtain that $\rho_0 \sim B^{1/2}$. 
The various system sizes fit in the same universal curve indicating that the
finite size effects are negligible.
Note that in the lattice case the density of states at low energies
is linear \cite{FT,Marinelli,PRB,AV} (this result differs from
the behavior obtained by others for a $d$-wave superconductor with no
disorder \cite{Volovik,Wang},
where $\rho(\epsilon \sim 0) \sim B^{1/2}$).
The finite density of states at zero energy is therefore a
consequence of finite disorder.

\begin{figure}
\includegraphics[width=7cm,height=7cm]{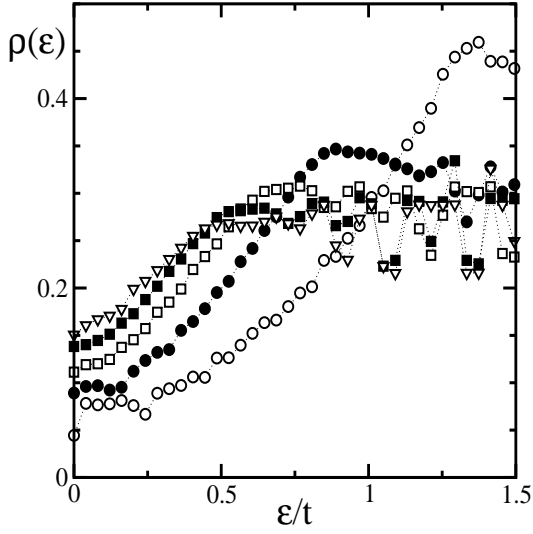}
\caption{\label{fig10}Quasiparticle density of states for the $d$-wave case
 for different values of the anisotropy parameter:
$\alpha_D=1$ ($\circ$),
$\alpha_D=2$ ($\bullet$),
$\alpha_D=3$ ($\square$), 
$\alpha_D=4$ ($\blacksquare$)
and 
$\alpha_D=5$ ($\nabla$). The magnetic field intensity is $B=1/100$.
The linear system size is $L=20$ and the chemical potential is $\mu=-2.2t$.}
\end{figure}

In Fig. \ref{fig9}
we focus on the narrow region close to $\epsilon=0$ for the
same set of parameters considered in Fig. \ref{fig8}.
Except for the lowest field case $B=1/200$ the
density of states seems to be finite at zero energy.
Here 
the field density $B=1/200$ corresponds to the particular
case where only two vortices pierced the $20\times20$ disordered supercell.
As shown in Ref. \onlinecite{Marinelli}
in this case the spectrum is usually gapped and therefore the density of
states vanishes at zero energy. Performing a fit like in Eq. \ref{powerlaw}
we obtain an
exponent which is now close to $1$.
In this regime $\rho_{0\rm{dip}}$
also scales with $\sqrt{B}$ and the slope scales linearly with $B$.
In this low energy regime the finite size effects are still noticeable
but the dependence on the magnetic field is common to the various
system sizes.
At these low energies the density of states for the various
system sizes appears to be of the following approximate form
\begin{equation}
\rho(\epsilon) \sim \frac{1}{\omega_H}\frac{1}{l^2}
{\cal F} \left( \frac{\epsilon}{\omega_H}
\frac{\delta^2}{l^2} \right),
\end{equation}
where $\omega_H \sim \sqrt{\Delta B}$ and ${\cal F}$
is
a universal function. In the left panel of Fig. \ref{fig9}
we show $\rho(\epsilon)$
for various fields while in the right panel
we illustrate the near scaling at low 
energies consistent with ${\cal F}(x) \sim c_1 + c_2 x$ at small $x$.
Note that $\beta (B\to 0)$ 
appears to be small but finite, consistent with a crossover to a
``Dirac node'' scaling 
$\rho(\epsilon) \sim (1/\omega_H)(1/l^2)
{\cal F} \left(\epsilon/\omega_H\right)$ at very low fields.

\subsection{Anisotropic case}

\begin{figure}
\includegraphics[width=8.5cm,height=8.5cm]{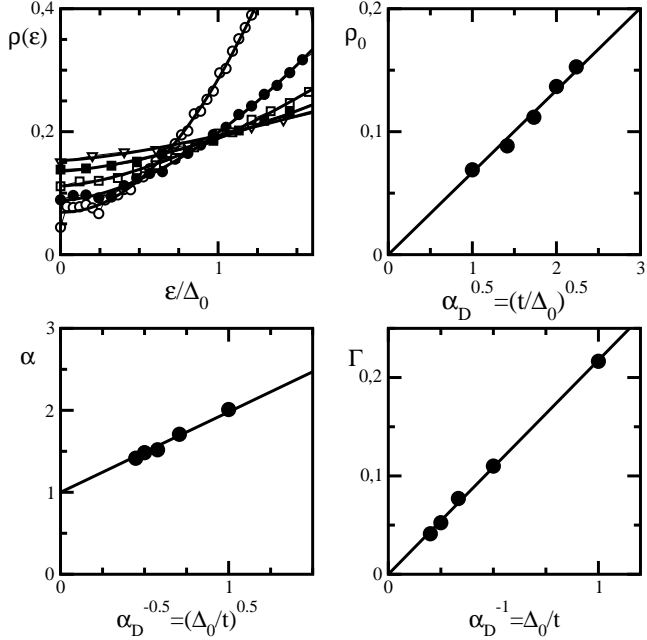}
\caption{\label{fig11}
Fits of the data presented in Fig. \ref{fig10}.
Upper left pannel:
quasiparticle density of states
presented in Fig. \ref{fig10} (the same symbols are used)
as a function of the rescaled energy $\epsilon/\Delta_0$.
Solid lines are fits of the power law type
$\rho(\epsilon)=\rho_0+\Gamma\left(\epsilon/\Delta_0\right)^\alpha$.
On the others pannel solid lines are linear fits.
Upper right pannel:
zero energy quasiparticle density of states $\rho_0$ as a function
of $\sqrt{\alpha_D}$.
Lower left pannel:
Power law exponent $\alpha$ as a function of $1/\sqrt{\alpha_D}$.
Lower right pannel:
Parameter $\Gamma$ of the power law as a function of $1/\alpha_D$.}
\end{figure}

We investigate now the dependence of the quasiparticle
density of state on the value of the Dirac anisotropy ratio $\alpha_D$.
Such a study is interesting in order to compare our results
with experiments; indeed for high-$T_c$ superconductors such as 
YBa$_2$Cu$_3$O$_7$ and Bi$_2$Sr$_2$CaCu$_2$O$_8$ 
the Dirac anisotropy ratio is\cite{Chiao1,Chiao2} respectively
$\alpha_D\simeq 14$ and $\alpha_D\simeq 19$.

\begin{figure}
\includegraphics[width=7cm,height=7cm]{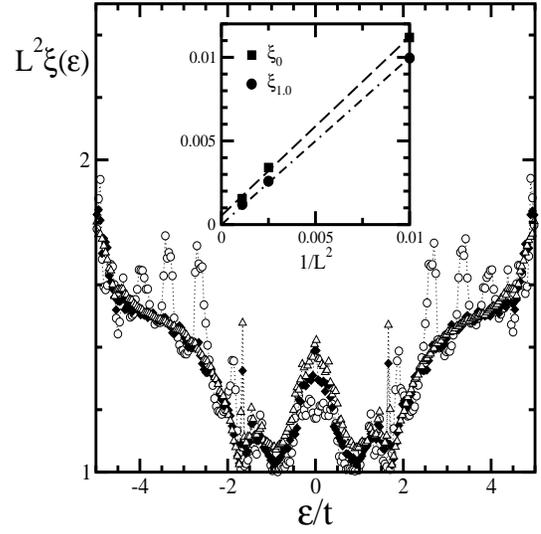}
\caption{\label{fig12}
Inverse participation ratio $\xi$ as a function of the quasiparticle energy
$\epsilon$ for $B=1/25$ and for different linear system sizes 
$L=10$ ($\circ$),
$L=20$ ($\blacklozenge$)
and
$L=30$ ($\triangle$).
Inset: scaling of the inverse participation ratio with $1/L^2$
for states in the close vicinity of the Fermi surface
$\left\arrowvert\epsilon\right\arrowvert\leq 0.05t$ ($\xi_0$, $\blacksquare$)
and for states with energy $\left\arrowvert\epsilon\right\arrowvert\simeq t$
($\xi_{1.0}$, $\bullet$).
}
\end{figure}

In the lattice case a high
anisotropy increases the density of states at low energies and leads to lines
of quasi-nodes \cite{PRB,Marinelli}. At high anisotropy ($\Delta_0<<t$)
the nodes
are very narrow and a one-dimensional like character is evidenced.

As shown in Fig. \ref{fig10}, the low-energy quasiparticle
density of states at constant field is filled when the anisotropy parameter 
$\alpha_D$ is increased. There is a narrow linear region close to the origin
that decreases as $\alpha_D$ increases.
Fig. \ref{fig11} shows the power law fits
of the data presented in Fig. \ref{fig10}
neglecting a very narrow region of width $\sim0.025t$
around $\epsilon=0$. 
It turns out that the low-energy quasiparticle density of states
has the form
\begin{equation}\label{rhoaniso}
\rho(\epsilon)\sim \rho_0+\Gamma\left(\frac{\epsilon}{\Delta_0}\right)^\alpha
\end{equation}
with zero-energy density of states
\begin{equation}
\rho_0\sim\sqrt{\frac{t}{\Delta_0}},
\end{equation}
the exponent
\begin{equation}
\alpha\simeq 1+\sqrt{\frac{\Delta_0}{t}}
\end{equation}
and the parameter
\begin{equation}
\Gamma\sim\frac{\Delta_0}{t}.
\end{equation}

\begin{figure*}
\includegraphics[width=14cm,height=14cm]{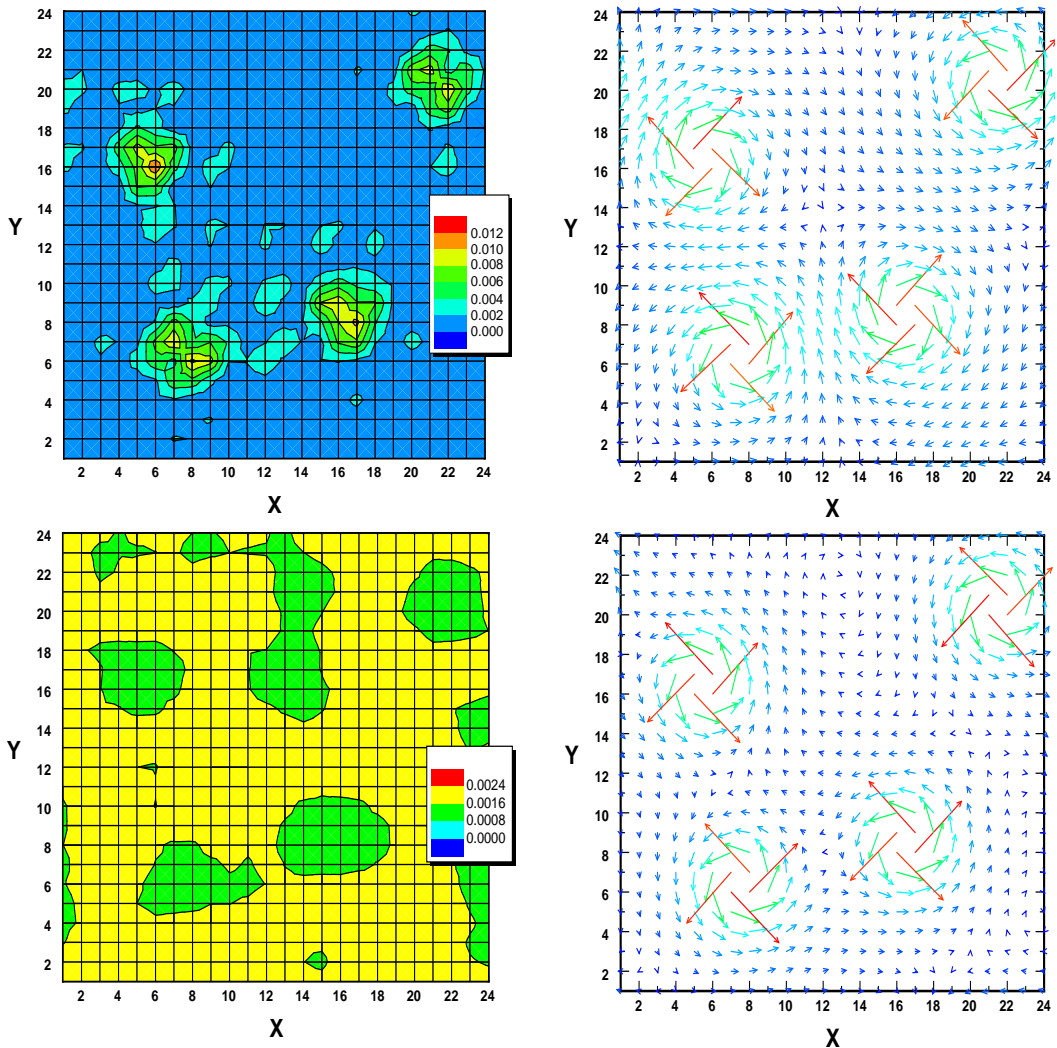}
\caption{\label{fig13}Upper left panel:
Color density
plot of the quasiparticle local density of states $\rho(\mbf{r},\epsilon)$
for $\epsilon\simeq0$.
Lower left panel:
Color density
plot of the quasiparticle local density of states $\rho(\mbf{r},\epsilon)$
for $\epsilon\simeq t$.
Upper right panel:
Color vector field plot showing the profile of the internal gauge field
$\mbf{a}_s=
\frac{1}{2}\left(\mbf{k}_s^A-\mbf{k}_s^B\right)$.
Lower right panel:
Color vector field plot showing the profile of the conventional
superfluid wave vector $\mbf{k}_s=
\frac{1}{2}\left(\mbf{k}_s^A+\mbf{k}_s^B\right)$.
All the panels correspond to the same configuration of the vortex pinning
and to the same magnetic field $B=1/144$.
}
\end{figure*}

The dependence of the exponent is interesting since in the isotropic
case we retrieve 
$\alpha=2$ and with increasing anisotropy the exponent decreases
to $1$ (see lower left panel of Fig. \ref{fig11})
characteristic of the Dirac nodes.
At $\epsilon\simeq0$ and for a high Dirac anisotropy ratio
the quasiparticle density of states
flattens since the coefficient $\Gamma$ in (\ref{rhoaniso}) decreases 
with $\Delta_0$.

\subsection{Spatial structure of the low-lying quasiparticle states}

As discussed above, the nature of the low lying states is an important point
which is relevant due to the presence of disorder. A standard way to analyse
the nature of the states is to study the IPR.
The IPR is defined in the usual way
\begin{equation}
\xi(\epsilon)=\frac{
\left\langle
\sum_{\mbf{n},\mbf{r}}
\left(
\left\arrowvert
u_{\mbf{n}}\left(\mbf{r}\right)
\right\arrowvert^4
+
\left\arrowvert
v_{\mbf{n}}\left(\mbf{r}\right)
\right\arrowvert^4
\right)\delta\left(\epsilon-E_{\mbf{n}}\right)
\right\rangle}
{
\left\langle
\sum_{\mbf{n},\mbf{r}}
\left(
\left\arrowvert
u_{\mbf{n}}\left(\mbf{r}\right)
\right\arrowvert^2
+
\left\arrowvert
v_{\mbf{n}}\left(\mbf{r}\right)
\right\arrowvert^2
\right)
\delta\left(\epsilon-E_{\mbf{n}}\right)
\right\rangle^2
}.
\end{equation}
The brackets denote the averaging over disorder configurations. The
IPR $\xi$ is a direct measure of the spatial extend
of the quasiparticle wavefunctions. It scales as $1/L^2$ for extended states
and is constant for localized states with localization length $\ell_c<L$.

Fig. \ref{fig12}
presents the IPR $\xi$ as a function of
the quasiparticle energy $\epsilon$ and for different system sizes $L$.
All the states are well extended, since there is no size dependence
for the quantity $L^2\xi(\epsilon)$,
except those in the close vicinity of
the Fermi surface ($\arrowvert\epsilon\arrowvert<0.5t$).
In the inset of Fig. \ref{fig12} we present
the scaling of the IPR $\xi$
with $1/L^2$. We consider an
average over states at the Fermi surface 
with $\arrowvert\epsilon\arrowvert<0.05t$ ($\xi_0$ in the inset)
and an average over states with $\epsilon\simeq t$ ($\xi_{1.0}$ in the inset).
The states close to $\epsilon=t$ are clearly extended since
the scaling with $1/L^2$ is quite accurate. The nature of the states close to
$\epsilon=0$ is however less clear. From Fig. \ref{fig12} we see
that the quasiparticle states close to $\epsilon=0$ deviate from the strict
$1/L^2$ scaling law (see the weak intercept of $\xi_0$ in the inset of Fig.
\ref{fig12}), however our results show an obvious size dependence 
for $\xi_0$ and do not show a saturation of $\xi_0$ with the linear system
size $L$. This latter fact indicates, in the hypothesis of an eventual
localization of these low-lying states, that we are still far from the
thermodynamic limit with the linear system sizes we can currently attain
within our model.

To gain further insight into the nature of the low lying states we show
in Fig. \ref{fig13} the LDOS for the states close to
$\epsilon=0$ and those close to $\epsilon=t$. At low energies
the LDOS defined by
\begin{equation}\label{ldos}
\rho(\mbf{r},\epsilon)=
\sum_{\mbf{n}}
\left(
\left\arrowvert
u_{\mbf{n}}(\mbf{r})
\right\arrowvert^2
+
\left\arrowvert
v_{\mbf{n}}(\mbf{r})
\right\arrowvert^2
\right)
\delta(\epsilon-E_{\mbf{n}})
\end{equation}
is strongly peaked near the vortex cores (upper left panel of Fig.
\ref{fig13}).
At higher energies ($\epsilon\simeq 1$)
the LDOS is much more homogeneously spread over the
system indicative of extended states; the values of the LDOS 
over the lattice are approximatively constant and
slightly fluctuate around the expected value
for extended states $1/L^2\simeq0.0017$ (lower left panel of Fig.
\ref{fig13}).

\begin{figure}
\includegraphics[width=8.5cm,height=6.5cm]{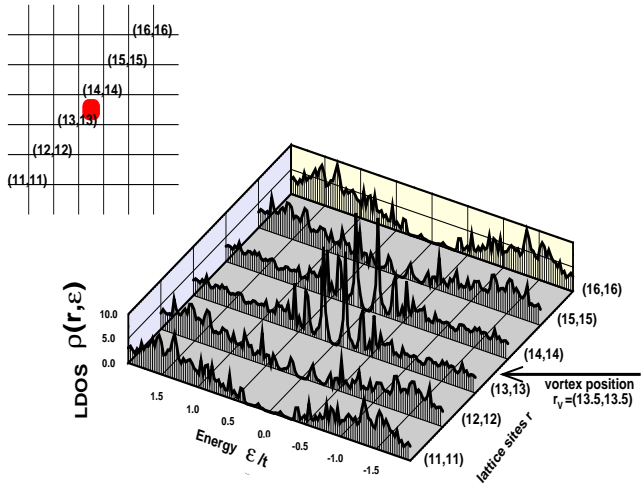}
\caption{\label{fig14}
Quasiparticle local density of states $\rho(\mbf{r},\epsilon)$
for the $d$-wave case at different lattice sites
$\mbf{r}=(11,11)$, $(12,12)$, $(13,13)$, $(14,14)$, $(15,15)$, $(16,16)$
around a vortex position $\mbf{r}_{\mbf{v}}=(13.5,13.5)$
belonging to a regular vortex lattice.
The magnetic field is $B=1/144$.
The other parameters are $\Delta_0=t$ and $\mu=-2.2t$.}
\end{figure}

The right side panels of Fig. \ref{fig13} show the profiles of the
internal gauge field $\mbf{a}_s$ (upper panel)
and of the conventional superfluid
wave vector $\mbf{k}_s$ (lower panel) corresponding to the same configuration
of vortex pinning as used in the computation of the LDOS
(left side panels). For the profile of the superfluid
wave vector $\mbf{k}_s$  the intensity of the supercurrents
is peaked around each vortices 
and rapidly decreases as the inverse of the 
square of the distance to the vortices.
The profile of the internal gauge field $\mbf{a}_s$ is more interesting.
It depicts the two types of vortices $A$ and $B$ whirling in opposite
directions (see upper right panel of Fig.
\ref{fig13}).
Due to this opposite circulation around $A$- and $B$-type vortices,
the interaction between a $A$- and a $B$-type vortex produces significant 
field currents between the vortices. Fig. \ref{fig13} shows a spatial
correlation
between these intervortex currents and the inhomogeneity of the local
density of states between the vortices for the low-lying states 
$\epsilon\simeq0$ (see upper left panel of Fig. \ref{fig13}).

If we compare now the results for the LDOS 
with those for the IPR $\xi$ for the
low-lying states, we can argue the following: as the participation ratio 
$\xi_0^{-1}$ physically counts the number of lattice sites occupied by the
quasiparticle wave function $\Psi(\mbf{r})$, and as the low energy states
are mainly located around the
vortices, the participation ratio $\xi_0^{-1}$ should scale
for a fixed magnetic field
as the number of vortices in a supercell $\xi_0^{-1}\sim N_\phi\sim BL^2$.
This argument explain the fact that $\xi_0$ in the inset of Fig.
\ref{fig12} seems to follow the $1/L^2$ without saturation, the weak intercept
being then negligible.
The low-lying quasiparticle states appear then to be delocalized
although strongly peaked around the vortex cores.

\begin{figure}
\includegraphics[width=8.5cm,height=6.5cm]{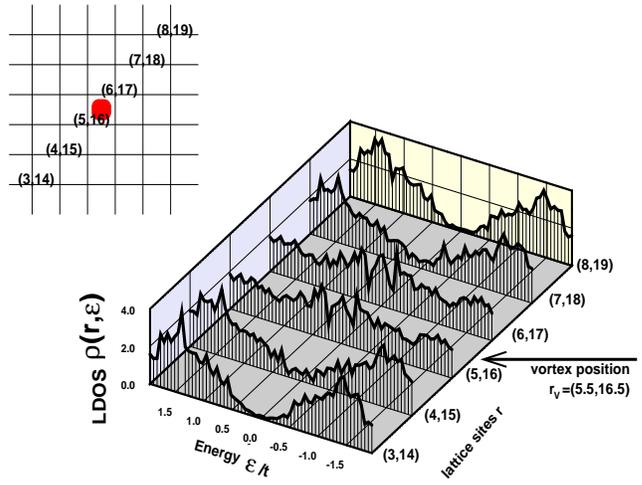}
\caption{\label{fig15}
Quasiparticle local density of states $\rho(\mbf{r},\epsilon)$
for the $d$-wave case at different lattice sites
$\mbf{r}=(3,14)$, $(4,15)$, $(5,16)$, $(6,17)$, $(7,18)$, $(8,19)$
around a vortex position $\mbf{r}_{\mbf{v}}=(5.5,16.5)$ belonging to
a disordered vortex lattice.
The magnetic field is $B=1/144$
and we use the 
$24\times24$ disordered supercell used in Fig. \ref{fig13}.
The other parameters are $\Delta_0=t$ and $\mu=-2.2t$.}
\end{figure}

We present now a spatial scanning of the LDOS
in the vicinity of a vortex core for the
case of a regular vortex lattice (Fig. \ref{fig14}) and 
for the case of a disordered vortex lattice (Fig. \ref{fig15}).
For both cases the LDOS away from the vortex
core are qualitatively the same and are comparable to that of a
$d$-wave superconductor in a zero-magnetic field.
Also for both cases the low-lying states are predominant
in the close vicinity of the vortex core 
but their respective spectra are different. 
For the regular vortex lattice case (Fig. \ref{fig14})
we remark a double peak structure around the vortex core.
This result is in qualitative agreement with the double peak
in conductance observed in YBa$_2$Cu$_3$O$_{7-\delta}$ by scanning
tunneling microscopy (STM) \cite{Maggio}. For the disordered vortex
case (Fig. \ref{fig15}) zero energy peaks (ZEP) appear in the close vicinity
of the vortex core (closest neighbor sites) and rapidily disappear
(typically over $3\delta$) when moving 
away from the vortex core. 
We note also 
that close to the vortex cores the coherence peaks dissapear in both
the regular and the disordered cases.

\section{Conclusion}

In summary, we have calculated the density of states of a disordered
superconductor in a pinned fully random vortex array. Both the disorder
and the magnetic field fill the density of states at low energies. 
In the $s$-wave case the density of states behaves as a power law with
an exponent that scales with $1/\sqrt{B}$. 
In general we
find a finite density
of states at zero energy for the $d$-wave case except in the limit of 
very small magnetic fields.
The density of states deviates from the zero energy value by a power law.
The zero energy density of states scales with the inverse of the magnetic
length ($\sqrt{B}$). 
In the $d$-wave case the Dirac anisotropy further increases the weight
of the density of states at low energies. Also it affects the exponent
of the power law. In the isotropic case the
exponent is $1$ at very low energies
and around $2$ neglecting this narrow region. In the linear regime the density
of states is of the form of a scaling function of the energy and the magnetic
field. As the anisotropy increases this narrow regime shrinks considerably and,
neglecting this region, the exponent of the density of states interpolates
to $1$ which is the Dirac limit. This limit is also obtained in the zero field
limit in the isotropic case.
Except for the zero energy finite value the energy
dependence of the density of states in the case 
with disorder is similar to the
lattice case.
This suggests that the vortex disorder does not dramatically affect the
density of states at low energies.
An analysis of the IPR and the LDOS shows that the lowest lying states
are delocalized, even though strongly peaked at the vortex cores.  
In the gapped $s$-wave case however
the disorder introduces states in the gap thereby changing
qualitatively the low energy density of states, as 
in the high field limit \cite{Sacramento}.
We found a power law behavior with an exponent that scales with the
magnetic length.

\begin{acknowledgments}
This work was supported in part by NSF grant DMR00-94981 (ZT)
and by FCT Fellowship SFRH/BPD/5602/2001 (JL).
\end{acknowledgments}


\begin{thebibliography}{99}
\bibitem{Gorkov} L. P. Gorkov and J. R. Schrieffer,
                 Phys. Rev. Lett. {\bf 80}, 3360 (1998).
\bibitem{Anderson} P. W. Anderson, cond-mat/9812063.
\bibitem{FT} M. Franz and Z. Te\v{s}anovi\'c,
             Phys. Rev. Lett. {\bf 84}, 554 (2000).
\bibitem{Volovik} G. E. Volovik,
                  Pis'ma Zh. \'Eksp. Teor. Fiz. {\bf 58}, 457 (1993);
                  [JETP Lett. {\bf 58}, 469 (1993)].
\bibitem{report} A. Altland, B. D. Simons, and M. R. Zirnbauer,
                 Phys. Rep. {\bf 359}, 283 (2002).
\bibitem{NTW} A. A. Nersesyan, A. M. Tsvelik, and F. Wenger,
              Phys. Rev. Lett. {\bf 72}, 2628 (1994);
              Nucl. Phys. B {\bf 438}, 561 (1995).
\bibitem{Senthil} T. Senthil, M. P. A. Fisher, L. Balents, and C. Nayak,
                  Phys. Rev. Lett. {\bf 81}, 4704 (1998).
\bibitem{did} R. B. Laughlin,
              Phys. Rev. Lett. {\bf 80}, 5188 (1998);
              T. Senthil, J. B. Marston, and M. P. A. Fisher,
              Phys. Rev. B {\bf 60}, 4245 (1999).
\bibitem{Dukan} S. Dukan and Z. Te\v{s}anovi\'c,
                Phys. Rev. B {\bf 56}, 838 (1997).
\bibitem{Sacramento} P. D. Sacramento,
                     Phys. Rev. B {\bf 59}, 8436 (1999).
\bibitem{Ye} J. Ye, Phys. Rev. Lett. {\bf 86}, 316 (2001).
\bibitem{Khveshchenko} D. V. Khveshchenko and A. G. Yashenkin, cond-mat/0204215.
\bibitem{prl} J. Lages, P. D. Sacramento, and Z. Te\v{s}anovi\'c, cond-mat/0301170.
\bibitem{Caroli} C. Caroli, P. G. de Gennes, and J. Matricon, Phys. Lett. A {\bf 9}, 307 (1964).
\bibitem{dvortex} M. Franz and Z. Te\v{s}anovi\'c, Phys. Rev. Lett. {\bf 80}, 4763 (1998).
\bibitem{PRB} O. Vafek, A. Melikyan, M. Franz, and Z. Te\v{s}anovi\'c,
              Phys. Rev. B {\bf 63}, 134509 (2001).
\bibitem{Chiao1} M. Chiao, R. W. Hill, C. Lupien, B. Popi\'c, R. Gagnon,
                and L. Taillefer, Phys. Rev. Lett. {\bf 82}, 2943 (1999).
\bibitem{Bessa} J. Lowell and J. B. Sousa, J. Low Temp. Phys. {\bf 3}, 65 (1970).
\bibitem{Atkinson} W. A. Atkinson, P. J. Hirschfeld and A. H. MacDonald, 
                   Phys. Rev. Lett. {\bf 85}, 3922 (2000).
\bibitem{Franzet} M. Franz, C. Kallin and A. J. Berlinsky, Phys. Rev. B {\bf 54}, R6897 (1996);
M. Franz et al. {\it ibid} {\bf 56}, 7882 (1997).
\bibitem{Randeria1} A. Ghosal, M. Randeria and N. Trivedi, Phys. Rev. B {\bf 63}, 020505 (2000).
\bibitem{Ulm} E. R. Ulm et al., Phys. Rev. B {\bf 51}, 9193 (1995);
D. N. Basov et al., Phys. Rev. B {\bf 49}, 12165 (1994);
C. Bernhard et al., Phys. Rev. Lett. {\bf 77}, 2304 (1996);
S. H. Moffat, R. A. Hughes, and J. S. Preston, Phys. Rev. B {\bf 55}, 14741 (1997).
\bibitem{Randeria2} A. Ghosal, M. Randeria, and N. Trivedi, Phys. Rev. Lett. {\bf 81}, 3940 (1998).
\bibitem{Inside} P. I. Soininen, C. Kallin, and A. J. Berlinsky, Phys. Rev. B {\bf 50}, 13883 (1994).
\bibitem{Marinelli} L. Marinelli, B. I Halperin, and S. H. Simon,
               Phys. Rev. B {\bf 62}, 3488 (2000).
\bibitem{AV} A. Vishwanath,
               Phys. Rev. B {\bf 66}, 064504 (2002).
\bibitem{Wang} Y. Wang and A. H. MacDonald,
               Phys. Rev. B {\bf 52}, R3876 (1995).
\bibitem{Chiao2} M. Chiao, R. W. Hill, C. Lupien, L. Taillefer, P. Lambert,
                 R. Gagnon, and P. Fournier,
                 Phys. Rev. B {\bf 62}, 3554 (2000).
\bibitem{Maggio} I. Maggio-Aprile, Ch. Renner, A. Erb, E. Walker, and
                 \O. Fischer, Phys. Rev. Lett. {\bf 75}, 2754 (1995).
\end{thebibliography}
\end{document}